\documentclass[10pt, conference, compsocconf]{IEEEtran}

\usepackage[cmex10]{amsmath}
\usepackage{graphicx}
\graphicspath{{./figures/}}
\DeclareGraphicsExtensions{.pdf,.png}
\usepackage{array}
\usepackage{amssymb} 
\usepackage{url}
\usepackage{xspace}
\usepackage{setspace}
\usepackage{balance}
\usepackage{cite}
\usepackage{pifont}
\newcommand{\cmark}{\ding{51}}%
\newcommand{\xmark}{\ding{55}}%
\usepackage{color}

\usepackage[skipabove=5,skipbelow=5,innertopmargin=5,innerbottommargin=5]{mdframed}

\newcommand{\ie}{{i.e.},\xspace}
\newcommand{\eg}{{e.g.},\xspace}

\newcommand{\etal}{et al.}

\newcommand{\GH}{{\sc GitHub}\xspace}
\newcommand{\GHT}{{\sc GHTorrent}\xspace}
\newcommand{\TR}{{\sc Travis-CI}\xspace}
\newcommand{\JK}{{\sc Jenkins}\xspace}

\begin{document}
%
\title{Continuous integration in a social-coding world: Empirical evidence from \GH\\
\emph{Updated version with corrections}\vspace{-0.8cm}}

\author{\IEEEauthorblockN{Bogdan Vasilescu, Stef van Schuylenburg, Jules Wulms, Alexander Serebrenik, Mark G. J. van den Brand}
\IEEEauthorblockA{Eindhoven University of Technology,\\ Den Dolech 2, P.O. Box 513, \\ 5600 MB Eindhoven, The Netherlands\\
\{b.n.vasilescu, a.serebrenik, m.g.j.v.d.brand\}@tue.nl, \{s.b.v.schuylenburg, j.j.h.m.wulms\}@student.tue.nl}
}


%


\maketitle


\begin{abstract}

Continuous integration is a software engineering practice of frequently merging 
all developer working copies with a shared main branch, \eg several times a day.

With the advent of \GH, a platform well known for its ``social coding'' features 
that aid collaboration and sharing, and currently the largest code host in the 
open source world, collaborative software development has never been more 
prominent.
In \GH development one can distinguish between two types of developer contributions
to a project: \emph{direct} ones, coming from a typically small group of developers 
with write access to the main project repository, and \emph{indirect} ones, coming 
from developers who fork the main repository, update their copies locally, and 
submit pull requests for review and merger.

In this paper we explore how \GH developers use continuous integration as well
as whether the contribution type (direct versus indirect) and different project 
characteristics (\eg main programming language, or project age) are associated 
with the success of the automatic builds.


\end{abstract}

\section{Introduction}
\label{sec:intro}

Continuous integration (CI) is a software engineering practice of frequently merging 
all developer working copies with a shared main branch~\cite{duvall2007continuous}, 
\eg several times a day, or with every commit.
CI is typically supported by build servers that verify each integration automatically, 
running unit tests and reporting the results back to the developers.
The concept, often attributed to Martin Fowler based on a 2000 blog 
entry~\cite{fowler2006continuous} but known as ``“synch-and-stabilize'' 
or ``nightly build'' already in 1997~\cite{cusumano1997how}, 
is a recommended best practice of agile software 
development methods like eXtreme Programming~\cite{beck2000extreme}.
This continuous application of quality control checks aims to speed up the 
development process and to ultimately improve software quality, by reducing the 
integration problems occurring between team members that develop software 
collaboratively~\cite{duvall2007continuous}.

CI as a quality control mechanism in distributed, collaborative contexts is common 
to both commercial (\eg Microsoft~\cite{cusumano2007extreme, miller2008hundred}) 
and open source software (OSS) development (\eg FreeBSD 
and Mozilla~\cite{holck2007continuous}).
However, CI is especially relevant to OSS projects, due to the difficulties typically 
associated with imposing structured processes in such projects and on their 
contributors~\cite{holck2007continuous}: requirements documents are
often lacking~\cite{mockus2002two} and project contributors are often 
volunteers~\cite{hars2001working}, typically geographically 
distributed~\cite{dempsey2002open}, and rarely motivated by working in a 
team~\cite{lakhani2005hackers}.

With the advent of social media in (OSS) software development, recent years 
have witnessed  many changes to how software is developed, and how developers 
collaborate, communicate, and learn~\cite{
Dabbish2012Social, Begel2013SNmeetsSD, VasilescuCSCW14}.
One such prominent change is the emergence of the pull-based development
model~\cite{barr2012cohesive, gousios2014exploratory}, made possible by the
distributed version control system Git, and made popular by the ``social coding''
platform \GH, currently the largest code host in the OSS world.
In this model one can distinguish between two types of developer contributions
to a project: \emph{direct} ones, coming from a typically small group of developers 
with write access to the main project repository, and \emph{indirect} ones, coming 
from developers who fork the main repository, update their copies locally, and 
submit pull requests for review and merger.

\GH's implementation of the pull-based development model enables anyone with
an account to submit changes to any repository with only a few clicks.
This represents an unprecedented low barrier to entry for potential contributors,
but it also impacts testing behavior~\cite{Dabbish2012Social, pham2013creating}.
For example, \GH project owners interviewed by Pham \etal~\cite{pham2013creating} 
reported scalability challenges when integrating (many) outside contributions, driving 
them towards automated tests.
Automated CI services, such as \TR\footnote{\url{https://travis-ci.com}}---integrated 
with \GH itself---or \JK\footnote{\url{http://jenkins-ci.org}}, facilitate this 
process by automating a number of steps: whenever a commit is recorded or a 
pull request is received, the contribution is merged automatically into a testing 
branch, the existing test suite is run, and the contribution author and project 
owner are notified of the results.
However, despite these potential benefits, CI services are reportedly underused
both on \GH~\cite{pham2013creating} as well as in OSS in 
general~\cite{deshpande2008continuous}.

In this paper we focus on \TR, arguably the most popular CI service on 
\GH.\footnote{See, \eg the blog entries 
\url{https://blog.codecentric.de/en/2012/05/travis-ci-or-how-continuous-integration-will-become-fun-again/}
and \url{https://blog.futurice.com/tech-pick-of-the-week-travis-ci}, acc.\ June 2014}
Specifically, we quantitatively explore to what extent \GH developers use the \TR 
service, as well as whether the contribution type (direct versus indirect) or project 
characteristics (\eg main programming language, or project age) are associated with 
the success of the automatic builds.

The remainder of this paper is organised as follows. 
After discussing the methodology followed in Section~\ref{sec:methods}, we present 
our results in Section~\ref{sec:results}. 
Then, we review threats to validity in Section~\ref{sec:threats}, and finally 
we conclude in Section~\ref{sec:conclusions}.

\section{Methodology}
\label{sec:methods}

To understand usage of the \TR service in \GH projects, we extracted and 
integrated data from two repositories: (i) \GHT~\cite{gousios2013ghtorent,Gousios2014Lean}, a 
service collecting and making available metadata for all public projects available 
on \GH; and (ii) the \TR API\footnote{\url{http://docs.travis-ci.com/api/}}.
This section describes how we collected and analysed these data.

\subsection{Sample Selection}

Due to limitations of querying the \TR API, we restricted our attention in this
study to a sample of large and active \GH projects.
Using the \GHT web interface\footnote{Accessible from \url{http://ghtorrent.org/dblite/}},
we selected all \GH repositories that: (i) are not forks of other repositories; (ii) have 
not been deleted; (iii) are at least one year old; (iv) receive both commits and pull 
requests; (v) have been developed in Java, Python or Ruby; (vi) had at least 10 
changes (commits or pull requests) during the last month; and (vii) have at least 10 
contributors. We choose projects that receive both commits and pull 
requests, since we want to understand whether the way modification has been submitted
(commit or pull request) can be associated with the build success. 
Our choice of the programming languages has been motivated by the 
history of \TR: \TR started as a service to the Ruby community in early 2011,
while support for Java and Python has been announced one year later (February 21, 2012
and February 27,  2012, respectively). We expect therefore the use of \TR to be more
widespread for Ruby than for Java and Python.

The data were extracted on March 30, 2014.

After filtering our sample contained 223 \GH projects, relatively balanced across
the three programming languages (Figure~\ref{fig:langs}): 70 (31.4\%) were coded
in Java, 83 (37.2\%) in Python, and 70 (31.4\%) in Ruby.
The sample includes many large and popular OSS projects, such as 
\texttt{rails}, \texttt{ruby}, 
\texttt{elasticsearch}, or \texttt{gradle}.

\subsection{Data Integration}

To extract data about the automatic builds, we started by querying the \texttt{repos} 
endpoint of the \TR JSON API (using the repository slug---username/repo---as argument), 
to determine whether \TR is configured for a particular project.
Then, if the response was not empty, we iteratively queried the builds associated 
with this project (25 at a time as per the \TR API) from the \texttt{builds} endpoint, 
collecting the \texttt{event\_type} fields (that distinguish pull requests from pushes)
and the \texttt{result} fields (that specify whether the build succeeded---0, or failed---1).
Ongoing builds, for which the \texttt{result} fields are not set, were ignored.

\subsection{Statistical Analysis}
\label{sec:stats}

We aggregated the data collected from the \TR API into contingency tables, one for
each \GH project, with rows corresponding to commits and pull requests, and 
columns---to passed and failed builds.
Then, to test whether the success/failure of the build is independent from the way
the modification has been proposed, we applied the $\chi^2$ test of 
independence. 
Next, to formalise the strength of this dependence, we calculated the odds
ratios 
and corresponding $p$-values.
Finally, to aggregate the results of the $\chi^2$ tests (one per project), we applied 
the Stouffer test using the weighted Z-score method~\cite{whitlock2005combining, 
stouffer1949american}.
This allows us to lift the results of the individual $\chi^2$ tests to the group level.


\section{Results}
\label{sec:results}

\subsection{Direct Versus Indirect Contributions}

We start by investigating the preference for direct (pushes) and indirect (pull requests)
contributions among the projects in our sample.
Java projects used the fewest pull requests during the observation month (March 2014), 
with a maximum of 26. 
Among the Java projects are also the most projects that do not use pull requests at all 
out of the 3 languages. 
Python and Ruby projects both have higher counts of pull requests, with maxima of 97 
and 236, respectively. 
Some Python and Ruby projects even have more pull requests than commits.

The shared repository model (with contributors having write access to the repository)
is more popular among Java projects, while Python and Ruby projects have more 
contributors submitting pull requests. 
Overall, we see that commits (direct code modifications) are more popular than pull 
requests (indirect code modifications), with only a small number of projects having 
more pull requests than commits.
Similar findings have been reported by Gousios, Pinzger, and van Deursen in their
exploratory study of the pull-based software development model on \GH~\cite{gousios2014exploratory}.

\begin{mdframed}
Direct code modifications (pushed commits) are more popular than indirect code 
modifications (pull requests).
\end{mdframed}

\subsection{Usage of \TR}
 
\begin{figure}[t]
\centering
\includegraphics[width = 0.9\columnwidth, clip=true, trim=12 20 12 12]{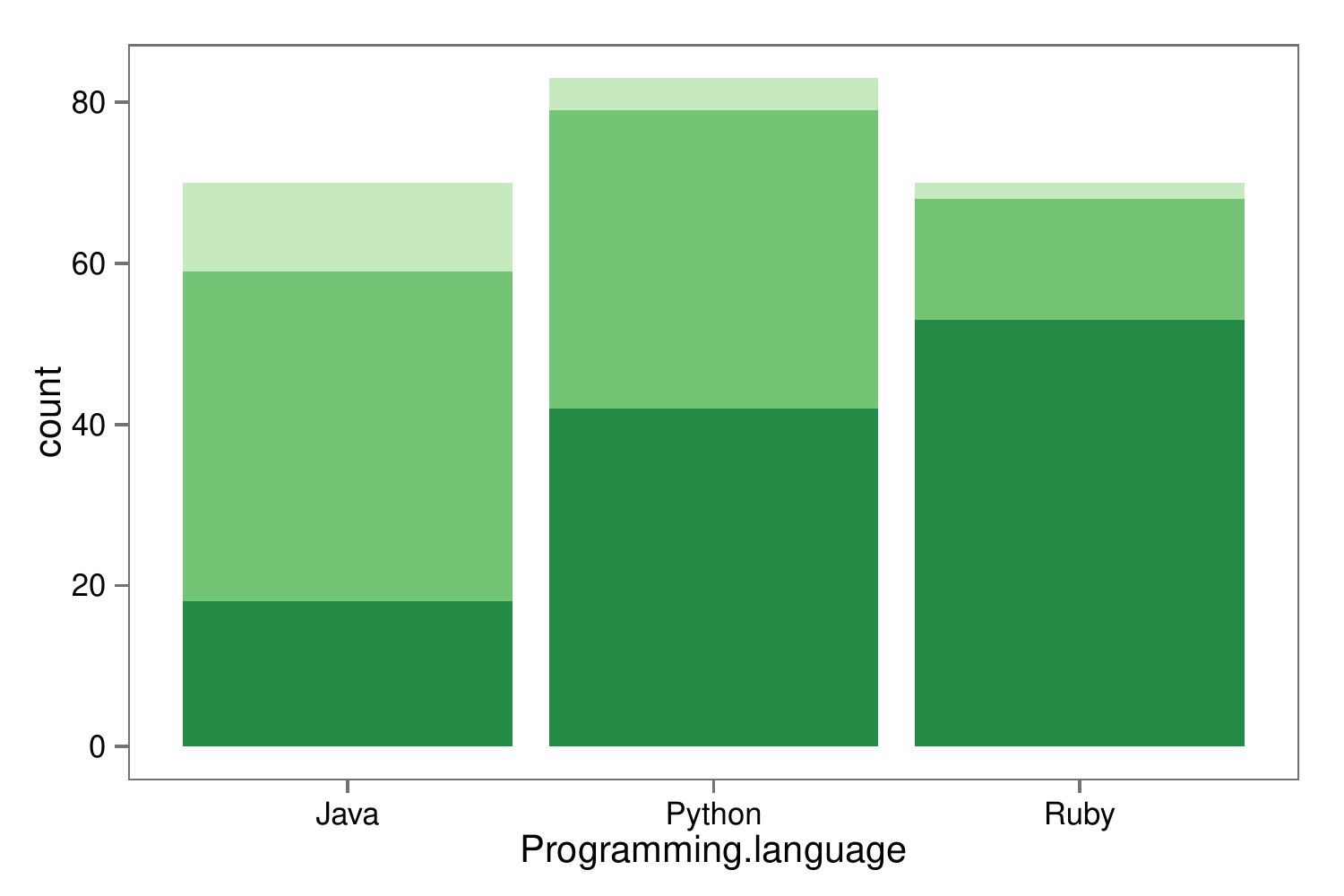}
\caption{Distribution of the \GH projects considered per programming language: 
while many Ruby projects are configured for \TR and use it (dark green), most Java projects
are configured for \TR but do not use it (middle green). Eleven Java projects are not
configured for \TR (light green).}
\label{fig:langs}
\end{figure}

Next we investigate usage of \TR among the projects in our sample.
First, we observe that an overwhelming majority of the projects are configured to use 
\TR (206 out of 223 projects, or 92.3\%), confirming the anecdotal popularity of the CI
service among \GH developers.
However, slightly less than half of the 206 projects (93, or 45\%) have no associated 
builds recorded in the \TR database. 
This shows that while most projects \emph{are ready} to use continuous integration, 
significantly fewer \emph{actually do}.
When we look into the distribution of these projects with no builds with respect to 
programming language, we observe that Java projects are overrepresented (69.5\%
of the Java projects configured to use \TR are not actually using it), while Ruby 
projects are underrepresented (Figure~\ref{fig:langs}).

\begin{mdframed}
Although most \GH projects in our sample are configured to use the \TR continuous 
integration service, less than half actually do. 
In terms of languages, Ruby projects are among the early adopters of \TR, while 
Java projects are late to use continuous integration.
\end{mdframed}

\subsection{Contribution Type and Build Success}
\label{sec:CTBS}
We have observed that the median success rate is 79.5\% for commits and  
75.9\% for pull requests. 
To better understand whether \TR build failure 
is independent from the way the modification has been proposed,
we focussed on projects with at least 5 failed/successful 
builds for each contribution type, as required by the $\chi^2$ test of independence
(cf.\ Section~\ref{sec:stats}).
Out of 113 \GH projects configured to use \TR and actually using it ($206-93$, cf.\ the 
discussion in the previous subsection), 84 projects had sufficient data for the $\chi^2$ test.
Among the remaining 29 projects to which the $\chi^2$ test could not be applied, in
most cases it was the failed pull requests cell that had insufficient data.
In other words, builds fail less frequently when contributions are submitted via pull requests. 
We believe this is because when a developer does not have commit rights and she 
suggests a change via a pull request, she will try harder to make sure the change is 
valid and it will not break the build. 
However, when instead a developer has commit rights, she can try out new things 
more freely, since she also has the power to reverse the change, if necessary. 

\begin{figure}[t]
\centering
\includegraphics[width=0.9\columnwidth, clip=true, trim=30 25 25 40]{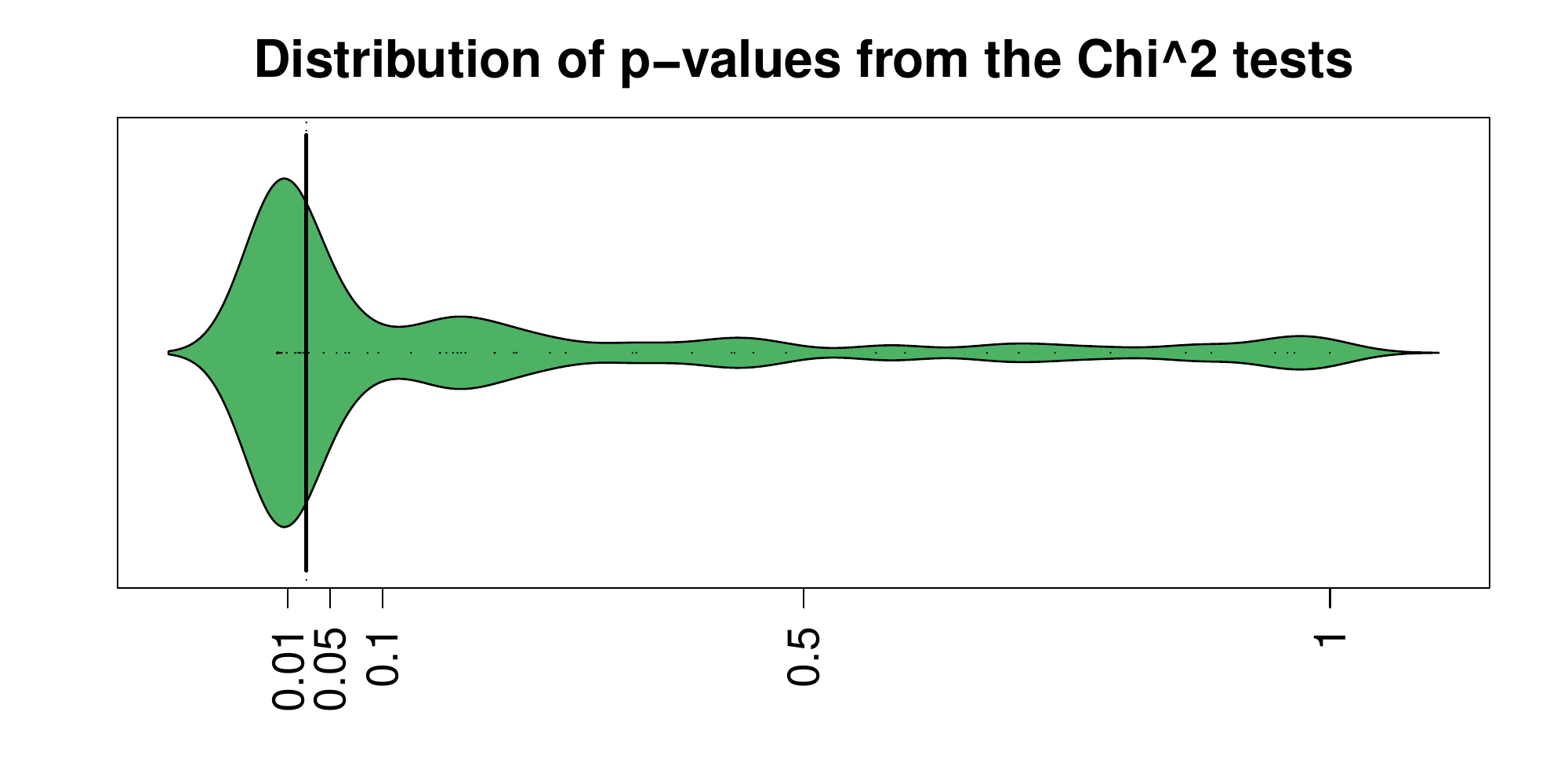}
\caption{Beanplot~\cite{kampstra2008beanplot} of the distribution of $\chi^2$ $p$-values
(84 \GH projects). The thick vertical line 
represents the median value.}
\label{fig:pvals}
\end{figure}

Finally, we lift the results of individual $\chi^2$ tests to the group level by applying the 
Stouffer procedure to obtain a combined significance level, \ie ``overall probability in 
favor of the outcome of the majority of the studies''~\cite{sheskin2007handbook}. 
The Stouffer test statistic $Z$ was calculated as $18.34$ and the corresponding 
$p$-value was too small to be calculated precisely (see Figure~\ref{fig:pvals} for 
the distribution of $p$-values of the individual $\chi^2$ tests).
This implies that taken together, the data indicates dependence of the build success 
on the way the modification has been proposed. 

\begin{figure}[t]
\centering
\includegraphics[width=0.9\columnwidth, clip=true, trim=30 45 25 40]{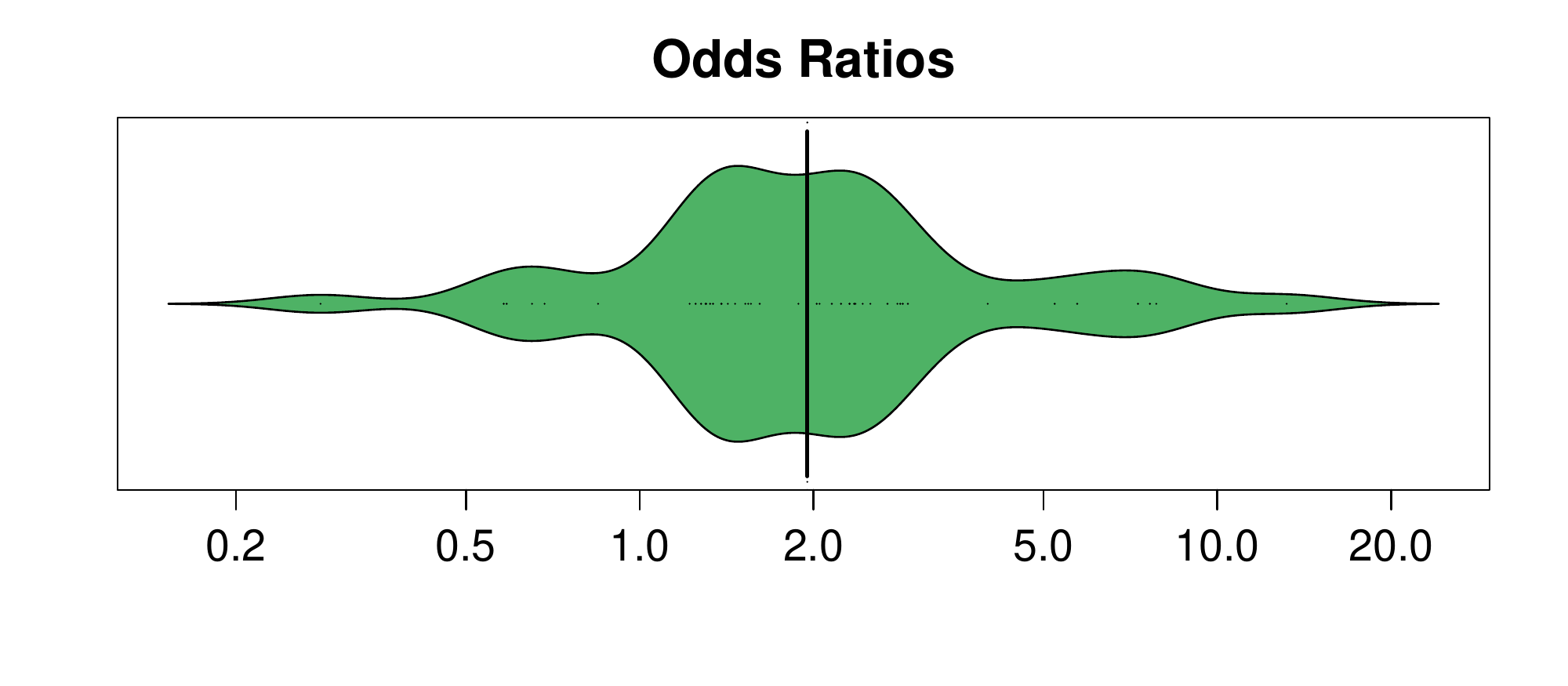}
\caption{Beanplot~\cite{kampstra2008beanplot} of the distribution of ratios between
the odds that commit builds succeed and the odds that pull request builds succeed
(45 \GH projects with statistically significant $\chi^2$ test results; $p < 0.05$). 
All odds ratios have $p < 0.05$. The thick vertical line shows 
the median.}
\label{fig:or}
\end{figure}

To investigate the directionality of this dependence we compute the odds ratios
(\ie the ratios between the odds that commit builds succeed and the odds that pull 
request builds succeed) for the \GH projects that showed statistically significant 
$\chi^2$ test results at 95\% confidence level.
Figure~\ref{fig:or} displays the distribution of the resulting odds ratios (all
statistically significant at 95\% confidence level).
{\color{red} Inspection of Figure~\ref{fig:or} reveals that for the overwhelming majority of 
projects (39 out of 45, or 87\%), builds corresponding to push commits are much 
more likely to succeed than builds corresponding to pull requests (since the
odds ratios are greater than 1).
This suggests that pull request evaluation is a complex process; 
pull requests are prone to integration testing failures, and other mechanisms
are needed to ensure quality control, \eg discussions and code 
review~\cite{gousios2014exploratory, tsay2014talk}.}


\begin{mdframed}
{\color{red}Pull requests are more likely to result in integration testing failures than push commits.}
\end{mdframed}

\subsection{Impact of Project Differences on the Contribution Type/Build Success Relation}

\begin{table}[b]
\footnotesize
\begin{tabular}{ l@{\hspace{0.1cm}}|l@{\hspace{0.1cm}}l@{\hspace{0.1cm}}l@{\hspace{0.05cm}}|l@{\hspace{0.1cm}}l@{\hspace{0.1cm}}l@{\hspace{0.02cm}}|l@{\hspace{0.1cm}}l@{\hspace{0.1cm}}l@{}}
& \multicolumn{3}{ c| }{Prog. lang.} & \multicolumn{3}{ c| }{Age (years)} & \multicolumn{3}{ c }{Contributors} \\
& Java & Python & Ruby & $<$2 & 2--4 & $>$4 & $\leq$17 & 17--33 & $>$33 \\ \hline
\# projects & 10 & 34 & 40 & 24 & 42 & 18 & 29 & 27 & 28 \\
\ldots s.t. $p<0.05$ & 3 & 19 & 23 & 9 & 25 & 11 & 18 & 15 & 12 \\ 
$H_0$ & \xmark & \cmark & \cmark & \xmark & \cmark & \cmark & \cmark & \cmark & \xmark\\
\%odds ratio$>$1 & n/a & 89 & 87 & n/a & 92 & 82 & 89 & 80 & n/a
\end{tabular}
\caption{Comparison of subgroups of 84 \GH projects based on the programming language, age and the number of contributors.}
\label{table:impact}
\end{table}

The 84 projects subjected to the $\chi^2$ test in Section~\ref{sec:CTBS} have been 
developed in different languages; have different ages; and involve different numbers 
of contributors. 
Table~\ref{table:impact} summarizes differences between those languages, ages, and 
numbers of contributors in terms of rejecting the null-hypotheses of the $\chi^2$ test, 
\ie independence of the build success from the way the modification has been proposed
(95\% confidence level). 
The thresholds of 17 and 33 contributors correspond to the 33\% and 67\% percentiles. 
Performing Stouffer tests for each group led to very small $p$-values, indicating that 
results can be lifted to the group level.
The data suggests that null hypotheses can be rejected (\cmark) for Python and Ruby 
projects, but not (\xmark) for Java projects; can be rejected for older 
projects but not for younger ones; and can be rejected for projects with not too many 
contributors as opposed to projects with many contributors.

Next we conducted odds ratio tests for projects where the null hypothesis has been 
rejected (\xmark) for the group level: {\color{red}all odds ratio tests turned out to be statistically
significant ($p$-values never exceeded 0.05) and in almost all cases the odds ratios 
exceeded 1. 
This means that whenever build success depends on the way the modification has 
been performed, pull requests are more likely to result in integration testing failures 
than direct commits.}


\begin{mdframed}
For Python and Ruby projects, projects older than two years, and projects with not too 
many contributors, {\color{red} pull requests are more likely to result in integration testing failures 
than direct commits.}
No such impact of the way the modification has been performed can be observed for 
Java projects, projects younger than two years, and projects with many contributors.
\end{mdframed}


\section{Threats to validity}
\label{sec:threats}

In this section, we discuss the threats to construct validity, internal validity 
and external validity~\cite{perry2000empirical}.

\emph{Construct} validity assesses whether the variables we considered accurately 
model our hypotheses. 
One of the threats to construct validity pertains to pull requests that 
appear as not being merged even if they have been merged~\cite{Kalliamvakou2014Promises}.

\emph{Internal validity} means that changes in dependent variables can be attributed 
to changes in independent variables instead of to something else. 
Specifically, we considered if the number of successful and failed builds
are related to the type of contribution
, the
main programming language, the project age, or the number of contributors.
We did not consider if other variables can confound this relationship. 
If such variables exist, they may invalidate our results.

\emph{External} validity means that the results we found can be generalized to 
real-world settings. 
Since we acquired a large set of data from a general-purpose website like 
\GH, we feel that our results can be generalized to large and active open-source
projects developed in Java, Python or Ruby and using \TR. 
Closed-source projects, small projects, projects in very different programming 
languages or using different CI services, such as \JK,
may not show the same patterns.


\section{Future work}
We plan to triangulate our quantitative findings through 
qualitative analysis, such as interviews and questionnaires. 
Conducting surveys can allow us to obtain insights in the ways
continuous integration is used in open-source and proprietary
development. 
A complementary approach will consist in performing a more 
detailed analysis of the \TR configuration files. 
Finally, we plan to consider a larger sample of \GH projects,
including those developed in additional programming languages.

\section{Conclusions}
\label{sec:conclusions}
In this paper we have studied a sample of large and active \GH projects 
developed in Java, Python and Ruby. 
We started by observing that direct code modifications (commits) are more 
popular than indirect code modifications (pull requests). 
Next, we have investigated the use of \TR: although most \GH projects in 
our sample are configured to use the \TR continuous integration service, 
less than half actually do. 
In terms of languages, Ruby projects are among the early adopters of \TR, 
while Java projects are late to use continuous integration. 
Next, for those projects that actually use \TR, we have studied whether the 
success or failure of a build is independent on the way code modification 
has been proposed. 
{\color{red}Our overall conclusion is that success or failure of a build does depend 
on the way the code modification has been proposed: pull requests are 
more likely to result in integration testing failures than direct commits.}
However, we observe differences for projects developed in different 
programming languages, of different ages, and involving different numbers 
of contributors.


\section*{Acknowledgements}

Special thanks to Mathias Meyer and the Travis CI team for helping us query 
their API.
Bogdan Vasilescu gratefully acknowledges support from the Dutch 
Science Foundation (NWO) through the NWO 600.065.120.10N235 project.

\balance

\bibliographystyle{IEEEtran}
\bibliography{references}

\end{document}